\documentclass{nst}
\usepackage{graphicx}
\usepackage{dcolumn}
\usepackage{bm}
\usepackage{longtable}
\usepackage{subfigure,dcolumn}
\usepackage{array}
\usepackage{booktabs}
\usepackage{verbatim}
\usepackage{etoolbox,xstring,mfirstuc,textcase}
\usepackage{amsmath,amsfonts,amssymb,amsthm} 
\usepackage{float}
\usepackage{tabularx}
\usepackage{tikz,xcolor,hyperref}
\definecolor{lime}{HTML}{A6CE39}
\DeclareRobustCommand{\orcidicon}{%
	\begin{tikzpicture}
	\draw[lime, fill=lime] (0,0) 
	circle [radius=0.16] 
	node[white] {{\fontfamily{qag}\selectfont \tiny ID}};	\draw[white, fill=white] (-0.0625,0.095) 
	circle [radius=0.007];	\end{tikzpicture}
	\hspace{-2mm}}
\foreach \x in {A, ..., Z}{%
	\expandafter\xdef\csname orcid\x\endcsname{\noexpand\href{https://orcid.org/\csname orcidauthor\x\endcsname}{\noexpand\orcidicon}}
	}

\usepackage{listings}
\begin{document}

\title{Commissioning of a radiofrequency quadrupole cooler-buncher for collinear laser spectroscopy} 

\thanks{This work was supported by the National Natural Science Foundation of China (Contract Nos.12027809, 12350007), National Key R\&D Program of China (Contract No.2022YFA1605100, 2023YFA1606403, 2023YFE0101600), New Cornerstone Science Foundation through the XPLORER PRIZE. The initial simulation and design studies for the MIRACLS RFQ cooler-buncher utilized and adapted in this work has received funding from the European Research Council (ERC) under the European Union’s Horizon 2020 research and innovation programme under grant agreement No. 679038.}

\author{Yin-Shen Liu}
\affiliation{School of Physics and State Key Laboratory of Nuclear Physics and Technology, Peking University, Beijing 100871, China}

\author{Han-Rui Hu}
\affiliation{School of Physics and State Key Laboratory of Nuclear Physics and Technology, Peking University, Beijing 100871, China}

\author{Xiao-Fei Yang\orcidB}
\email[Corresponding author ]{xiaofei.yang@pku.edu.cn}
\affiliation{School of Physics and State Key Laboratory of Nuclear Physics and Technology, Peking University, Beijing 100871, China}

\author{Wen-Cong Mei}
\affiliation{School of Physics and State Key Laboratory of Nuclear Physics and Technology, Peking University, Beijing 100871, China}

\author{Yang-Fan Guo}
\affiliation{School of Physics and State Key Laboratory of Nuclear Physics and Technology, Peking University, Beijing 100871, China}

\author{Zhou Yan}
\affiliation{School of Physics and State Key Laboratory of Nuclear Physics and Technology, Peking University, Beijing 100871, China}

\author{Shao-Jie Chen}
\affiliation{School of Physics and State Key Laboratory of Nuclear Physics and Technology, Peking University, Beijing 100871, China}

\author{Shi-wei Bai\orcidH}
\affiliation{School of Physics and State Key Laboratory of Nuclear Physics and Technology, Peking University, Beijing 100871, China}

\author{Shu-Jing Wang}
\affiliation{School of Physics and State Key Laboratory of Nuclear Physics and Technology, Peking University, Beijing 100871, China}

\author{Yong-Chao Liu}
\affiliation{School of Physics and State Key Laboratory of Nuclear Physics and Technology, Peking University, Beijing 100871, China}

\author{Peng Zhang}
\affiliation{School of Physics and State Key Laboratory of Nuclear Physics and Technology, Peking University, Beijing 100871, China}

\author{Dong-Yang Chen}
\affiliation{School of Physics and State Key Laboratory of Nuclear Physics and Technology, Peking University, Beijing 100871, China}

\author{Yan-Lin Ye\orcidD}
\affiliation{School of Physics and State Key Laboratory of Nuclear Physics and Technology, Peking University, Beijing 100871, China}

\author{Qi-Te Li}
\affiliation{School of Physics and State Key Laboratory of Nuclear Physics and Technology, Peking University, Beijing 100871, China}

\author{Jie Yang}
\affiliation{Institute of Modern Physics, Chinese Academy of Sciences, Lanzhou 730000, China}

\author{Stephan Malbrunot-Ettenauer\orcidF}
\affiliation{Experimental Physics Department, CERN, CH-1211 Geneva 23, Switzerland}

\author{Simon Lechner\orcidE}
\affiliation{Experimental Physics Department, CERN, CH-1211 Geneva 23, Switzerland}

\author{Carina Kanitz\orcidG}
\affiliation{Experimental Physics Department, CERN, CH-1211 Geneva 23, Switzerland}

\begin{abstract}

A RadioFrequency Quadrupole (RFQ) cooler-buncher system has been developed and implemented in a collinear laser spectroscopy setup. This system is dedicated to convert a continuous ion beam into short bunches, while enhancing beam quality and reducing energy spread. The functionality of the RFQ cooler-buncher has been verified through offline tests with stable rubidium and indium beam, delivered from a surface ion source and a laser ablation ion source, respectively. With a transmission efficiency exceeding 60\%, bunched ion beams with a full width at half maximum of approximately 2~$\mu$s in the time-of-flight spectrum have been successfully achieved. The implementation of the RFQ cooler-buncher system also significantly improves the overall transmission efficiency of the collinear laser spectroscopy setup.

\end{abstract}


\maketitle

\section{Introduction}

The fundamental properties of atomic nuclei, such as masses, nuclear spins, electromagnetic moments, and charge radii, are crucial for exploring exotic nuclear structures and probing the underlying nucleon-nucleon interactions~\cite{Ye2025,YANG2023104005,YAMAGUCHI2021103882,Deng2024,chenying2023,Wan2025}.
Laser spectroscopy techniques enable the precise determination of the electromagnetic properties of ground and long-lived isomeric states of atomic nuclei by measuring the hyperfine structure (HFS) spectra of their atoms, ions, and even molecules. Collinear laser spectroscopy (CLS)~\cite{YANG2023104005,Blaum_2013} is one such technique capable of achieving high-resolution HFS spectrum measurements. This is realized by overlapping a fast ion beam ($\sim$30 keV) with lasers in either collinear or anti-collinear geometry, effectively suppressing the spectral Doppler broadening caused by the energy spread of the ion beam~\cite{KAUFMAN1976309}.

There are two typical approaches used in CLS to measure the HFS spectrum. The most commonly used one is laser-induced fluorescence (LIF), which employs a continuous-wave narrow-band laser to excite atoms or ions from their ground or metastable states to higher excited states, followed by the detection of emitted fluorescence using photomultiplier tubes~\cite{Nörtershäuser2010}. However, the experimental sensitivity of this approach is often limited by high background signals caused by stray laser light. This limitation could be tamed by delivering the ion beam in short bunches, which enables to gate data taking on the ion-bunch passage through the laser-beam interaction region. As a result, the signal-to-background ratio is improved by 3-4 orders of magnitude, as first demonstrated in Ref. \cite{PhysRevLett.88.094801}. Another approach for measuring the HFS spectrum using CLS is resonance ionization spectroscopy (RIS), which utilizes multiple lasers to stepwise excite and then ionize the targeted atoms~\cite{VERNON2020384}. By detecting the resonantly laser-ionized ions with high efficiency, this approach eliminates the need for photon detection, thereby further improving the overall sensitivity of CLS~\cite{PhysRevC.96.041302}. In order to achieve high resonance ionization efficiency in this approach, high-power pulsed lasers are indispensable. Consequently, the ion beam must be delivered in the bunched mode to ensure proper temporal matching between the laser pulses and the ion beam bunches in the interaction region. An earlier attempt to use continuous ion beams for RIS in CLS resulted in limited efficiency due to the duty cycle losses~\cite{ALKHAZOV1991400}. Therefore, a bunched ion beam is a precondition for high-resolution and high-sensitivity HFS spectrum measurements using CLS.

\begin{figure*}
\begin{center}
\includegraphics[width=0.9\textwidth]{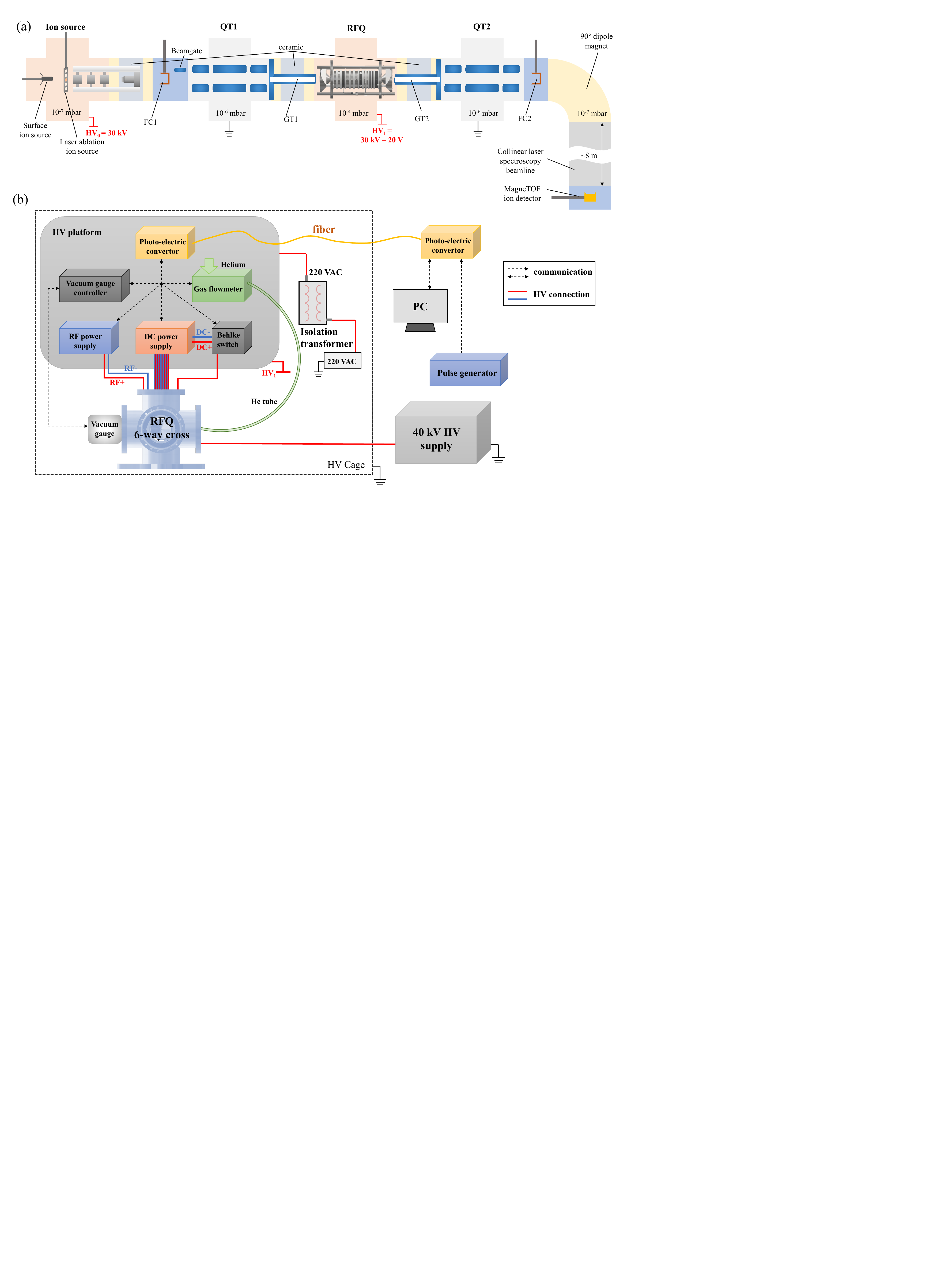}   
\end{center}
\vspace{-5mm}
\caption{
(a) Schematic of the RFQ cooler-buncher system, including the offline surface and laser ablation ion source, ion optics and the RFQ cooler-buncher. QT: Quadrupole Triplet; GT: Grounding Tube; FC: Faraday Cup.
(b) Layout of overall control and HV system. See text for more details.
}
\label{fig2}
\end{figure*}

In recent years, we have been working on the development of a CLS setup using both LIF and RIS approaches at the Radioactive Ion Beam (RIB) facilities in China \cite{BRIF, BRIF2,HIAF,HIRFL}. The first stage of the setup, based on the LIF approach, has already been implemented at the Beijing Radioactive Ion Facility (BRIF) of the China Institute of Atomic Energy. The first commissioning experiment successfully measured the HFS spectrum of the radioactive \(^{38}\text{K}\) isotope~\cite{WANG2022166622}.
However, compared to the $\sim$ 50~MHz linewidth of \(^{38}\text{K}\) HFS spectrum measured by the COLLAPS experiment at ISOLDE~\cite{Papuga:2112041}, the 150~MHz linewidth of HFS spectrum achieved at BRIF~\cite{WANG2022166622} is significantly broader.
This spectral broadening was found to originate from the considerable energy spread \((\delta E)\) of the ion beam delivered from the BRIF facility~\cite{WANG2022166622}. It was evaluated based on the Doppler broadening formula: 
\begin{equation}
  \Gamma_{\rm D} = \frac{\widetilde\nu_0{\delta E}}{\sqrt{2Emc^2}},
\end{equation}
where $\widetilde\nu_0$ = 25293.9~cm$^{-1}$ is the used laser wavenumber corresponding to the D1 line of potassium atomic transition. \\

The RadioFrequency Quadrupole (RFQ) cooler-buncher is a beam manipulation device designed to simultaneously satisfy the aforementioned beam requirements for CLS experiments, namely a pulsed ion beam with a low energy spread. This technique has been extensively implemented in the worldwide RIB facilities, such as IGISOL~\cite{PhysRevLett.88.094801}, ISOLDE~\cite{Mané:1198225}, NSCL~\cite{MINAMISONO201385}, and TRIUMF~\cite{BARQUEST2016207,Mané2011}, where it has proven effective in delivering high-quality ion beams for a range of applications.

In this work, we present a detailed commissioning test of a newly-installed RFQ cooler-buncher system, which is constructed based on a design~\cite{kanitz2019construction,MIRACLS_master} by the MIRACLS collaboration~\cite{MAIER2023167927} at ISOLDE-CERN. This system is tested by using a stable Rb ion beam in continuous mode from a surface ion source, and a stable In ion beam in bunched mode (with a typical bunch width of 100 $\mu$s) from a laser ablation ion source. Through a systemic test, the optimal operational parameters of the RFQ cooler-buncher system for the subsequent CLS experiment are identified. Under this condition, this system is capable of providing an ion bunch with a 2-$\mu$s temporal width, while maintaining 60\% overall transmission efficiency through the RFQ system.

\textbf{}

\section{RFQ cooler-buncher system}\label{sec:II}

Fig.~\ref{fig2}(a) presents a detailed schematic of the system, including the offline surface ion source and laser ablation ion source, ion optics, the RFQ cooler-buncher, as well as its control and HV platform. Monovalent positive ions are produced by the ion source and accelerated to $\sim$30 keV~\cite{PKU-CLS-2022}. The ion beam is reshaped by the electrostatic quadrupole triplet (QT1) lens for a better beam profile and transmission before injection into the RFQ cooler-buncher. The cooler-buncher is installed inside a six-way cross at a potential slightly lower than the 30-keV ion beam energy. Inside the cooler-buncher, the ions collide with the buffer gas atoms to eventually reach thermal equilibrium. In this manner, any potential large energy spread of the incoming ion beam is reduced. Simultaneously, the combined radial RF and axial DC electric fields enable trapping and accumulation of ions, thereby converting the incoming continuous ion beam into ion bunches with low (longitudinal and transverse) emittance, once released from the cooler-buncher. Following the extraction, the bunched ion beam is re-accelerated to 30 keV, and its beam profile is optimized  by the QT optics, prior to being directed into the CLS beamline. Two Faraday cups (FCs) are installed both upstream and downstream of the RFQ cooler-buncher to evaluate the ion beam transmission. The time structure, namely the time of flight (TOF) spectrum of the bunched ion beam is recorded by a MagneTOF ion detector and data acquisition (DAQ) system~\cite{CLS-DAQ-PKU}.

\subsection{Ion source}\label{sec:II-A}

Two types of offline ion sources are used for the RFQ system test: a newly installed surface ion source that provides a stable ion beam in continuous mode, and a laser ablation ion source that offers a stable ion beam in bunched mode. The surface ion source, containing Rb atoms (HeatWave Labs, \#101139), is mounted in a CF150 flange of a six-way cross, as shown in Fig.~\ref{fig2}(a). 
This ion source produces stable ion beams of $^{85}\text{Rb}$ (72.17\%) and $^{87}\text{Rb}$ (27.83\%). With a heating current of 1.44 A, an ion beam current of approximately 100~pA is measured at FC1 by optimizing the extraction electrode and lens in the ion source chamber, as shown in Fig.~\ref{fig2}.
It is worth noting that under these conditions, the average fluctuation of ion beam intensity remains within 1-pA for several hours, which is crucial for the RFQ cooler-buncher test. Details of the laser ablation ion source can be found in Ref.~\cite{PKU-CLS-2022}. 
In brief, an indium target is ablated by a 100 Hz 532-nm Nd:YAG laser (Beamtech Gama-M100) with a power density of 0.2 J/cm$^2$ and a pulse width of 10~ns, generating bunched stable indium ion beams with a temporal width of approximately 100 $\mu$s.

\subsection{Ion optics and beam diagnose}\label{sec:II-B}
As shown in Fig.~\ref{fig2} (a), the stable ion beam, after being extracted from the ion source, is accelerated to $\sim$30~keV and monitored directly by the FC1. A voltage of -50 V is applied to a grating in front of the cup to suppress secondary electron emissions when the ion beam impinges on the FC, ensuring accurate beam current reading at the level of 0.1 pA. 
The deflector installed after the FC1 serves as an equivalent beam gate. The voltage applied to the deflector (beamgate) is periodically modulated by a HV switch (Behlke Switch GHTS 30) controlled by a TTL signal. In this way, the beam current injected into the RFQ cooler-buncher can be reduced to below 1 pA, as the ion beam is deflected whenever the beamgate voltage is active.
This is to avoid potential space charge effects inside the RFQ cooler-buncher caused by the intense ion beam.

The QT1 lens upstream of the RFQ cooler-buncher is designed to ensure efficient transmission during the subsequent injection and deceleration process. A grounding tube (GT1), 150~mm in length and with an inner diameter of 20~mm, is installed inside a ceramic insulator just after the QT1, which connects the beamline on the ground potential with the floated vacuum chamber of the RFQ cooler-buncher. The purpose of GT1 is to provide a well-defined ground potential for the ions within the insulator tube, ensuring their trajectory is well controlled before they are injected into the RFQ cooler-buncher.. Additionally, this GT1 also functions as a gas-flow restricting tube, separating the $\sim$10$^{-6}$ mbar vacuum environment in the QT1 region and the $\sim$10$^{-4}$~mbar vacuum inside the RFQ cooler-buncher six-way cross. The GT2, QT2, and FC2, located downstream of the RFQ cooler-buncher (Fig.~\ref{fig2}(a)), have identical designs to those upstream, ensuring optimization of beam quality and evaluation of beam transport. All mentioned ion optics and FC are powered by $\pm$6-kV HV modules (EHS F0 60n/p) installed in an HV crate (ECH238, iseg company).

\begin{figure*}
\begin{center}
\includegraphics[width=0.9\textwidth]{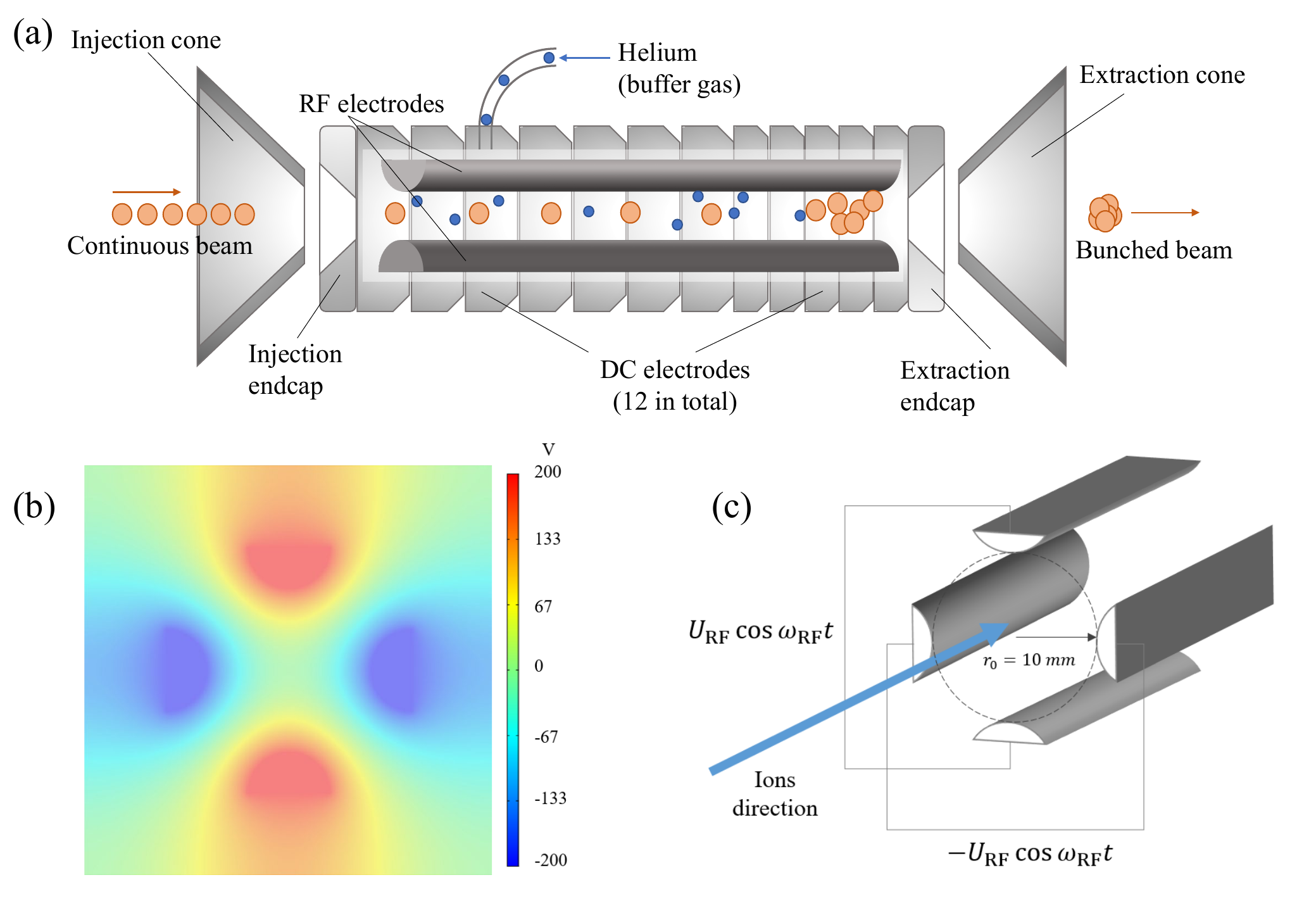} 
\end{center}
\vspace{-5mm}
\caption{{(a) A schematic diagram depicting the internal structure of the RFQ cooler-buncher. 
(b) Simulated potential profile of the RF field. (c) Voltages applied to the RF electrodes. Opposite electrodes are driven with voltages of the same polarity, while adjacent electrodes are driven with voltages of opposite polarity. This configuration creates a periodic RF field that confines the radial motion of ions.}}
\label{fig3}
\end{figure*}

\subsection{RFQ cooler-buncher}\label{sec:II-C}
A comprehensive description about the initial RFQ cooler-buncher’s design at MIRACLS will be presented in a forthcoming publication. Here, we describe its implementation and operation as part of the present work. As shown in Fig.~\ref{fig2}, the entire RFQ cooler-buncher is floated above a high potential of HV$_1$, isolated by two ceramic insulators. The internal structure of the cooler-buncher is shown in Fig.~\ref{fig3}(a). The injection section consists of two electrodes: the injection cone and the injection endcap. The potential of the injection cone is about HV$_1 - 2400$~V, while the injection endcap is set to around HV$_1 + 13$~V. Both voltages are supplied by the $\pm$6-kV HV modules (iseg, EHS 80 60n/p) inside a compact HV crate (iseg, ECH224) located on the HV platform at HV$_1$ (Fig.~\ref{fig2}(b)). The primary function of these two electrodes, in conjunction with the preceding GT1, is to gradually decelerate the ion beam and guide the ions into the RFQ cooler-buncher via a cone-shaped structure, and detailed simulations of the injection optics can be found in Refs.~\cite{kanitz2019construction,LECHNER2024169471}. Furthermore, the injection endcap features a minimum aperture with a diameter of 5~mm, thus ensuring a vacuum differential between the inside ($\sim$10$^{-2}$ mbar) and outside ($\sim$10$^{-4}$ mbar) of the RFQ cooler-buncher~\cite{LECHNER2024169471}.

Upon entering the cooler-buncher, ions are radially confined by the RF quadrupole electric field (Fig.~\ref{fig3}(b)) generated by RF electrodes, ensuring optimal ion transmission along the axial direction and trapping with minimal loss. The RF electrodes are composed of four identical semi-cylindrical rods, each with a radius of 5~mm and a length of 152 mm, with a minimum separation of 20~mm between opposing rods, as shown in Fig.~\ref{fig3}(c). The RF power supply (BGTPAX2231, designed and manufactured by Beijing BBEF Science \& Technology Co., Ltd.), provides two adjustable sine wave outputs with a frequency range of $f = 0.3 - 1.5$ MHz and an amplitude of $U_{\rm{RF}} = 0 - 200$V. These two identical RF signals, except for a 180$^\circ$ phase difference, are applied to the two pairs of opposing RF electrodes. Adjacent electrodes are driven 180$^\circ$ out of phase, while opposing electrodes share the same phase, thereby generating an RF field in the cross-sectional plane of the RFQ electrodes, as illustrated in Fig.~\ref{fig3}(b). The effective confinement of the ions' radial motion in the RF field is governed by the Mathieu equation~\cite{major2005charged}, which ensures stability when the parameter $0 \leq q \leq 0.908$, defined as:
\begin{equation}
q = \frac{e U_{\rm{RF}}}{m \pi^2 f^2 r_0^2}.
\end{equation}
Here, \( r_0 = 10 \) mm represents half the distance between the opposing RF electrodes.

The energy spread of ions can be reduced through collisions with the buffer gas atoms inside the cooler-buncher during trapping. In our system, helium with a purity of 99.999\% is chosen as the buffer gas. The flow rate of the injected helium gas is adjusted by a flowmeter (Sevenstar, CS200) with a maximum capacity of 100 Standard Cubic Centimeter per Minute (SCCM), and the gas is introduced into the cooler-buncher via a gas feedthrough on the top flange of the six-way cross. As ions traverse the cooler-buncher, they also experience continuous collisions with helium atoms. After a brief period (typically 1 to 5 ms, depending on the ion mass and gas flowrate~\cite{kanitz2019construction}), ions eventually reach thermal equilibrium with the helium atoms. 

Decelerated ions inside the cooler-buncher are also influenced by a DC electric field along the axial direction. This electric field, generated by 12 annular DC electrodes positioned along the axial direction, guides the ion beam towards the end section.
The first seven electrodes have a thickness of 13 mm, whereas the last five electrodes are 8 mm, allowing for more precise electric field control. The potentials of these 12 DC electrodes are supplied by ±500 V modules (iseg, EBS C0 05), forming an axial electric field together with the extraction endcap electrode. The extraction electrodes mirror those of the injection section; however the extraction endcap is not directly connected to the HV module. Instead, it interfaces with the HV module via a Behlke Switch (GHTS 30), which is controlled by a periodic TTL signal. The Behlke Switch features two HV inputs (DC+ and DC-) from $\pm$~6-kV HV modules (iseg, EHS 80 60n/p) and a single HV output, enabling the application of a periodic voltage to the extraction endcap. During a typical measurement cycle, the control TTL signal for the Behlke Switch remains at a low level, causing it to output a positive DC voltage (\textit{e.g.} +1600 V), thereby maintaining a high potential at the end section. Under the combined effects of the RF field, collisions with the buffer gas and DC trapping potential, ions are gradually cooled and accumulated at the end section. After enough trapping time, the control TTL signal switches to a high level, and the extraction endcap receives a negative voltage (\textit{e.g.} -800 V) from the Behlke Switch. Note that the voltage applied to the extraction cone is kept to a relatively large negative value (\textit{e.g.} -2000 V) for the efficient extraction. As a result, the ions trapped at the end section are released and re-accelerated to nearly 30 keV, eventually forming an ion bunch with a temporal width of only a few microseconds.

To achieve the desired functionality of the RFQ cooler-buncher, multiple hardware components are required, including an RF power supply, the helium gas input, DC power supplies and others. All devices associated with this system are illustrated in Fig.~\ref{fig2}(b). The entire cooler-buncher is floated at a high-voltage potential, HV$_{1}$, supplied by a high-precision DC power supply (Heinzinger PNChp 40000-15pos). Consequently, all associated devices are installed on an HV platform at the same potential, HV$_{1}$, within an HV cage, as shown in Fig.~\ref{fig2}(b). A 50-kV isolation transformer is employed to electrically isolate HV$_{1}$ from ground potential while supplying 220 VAC power to all the devices on the HV platform. To facilitate remote operation, all devices on the platform are controlled via pairs of photo-electric converters and optical fibers, ensuring bidirectional communication between the control PC and the equipment. This system allows for the control and/or monitoring of key components, such as the RF power supply, DC power supply, gas flowmeter, vacuum gauge controller and the Behlke Switch. The TTL signals required to operate the Behlke Switch are generated by a pulse generator (Quantum Composer 9520).

\section{Performance Test}

\begin{figure}
\begin{center}
\includegraphics[width=\linewidth]{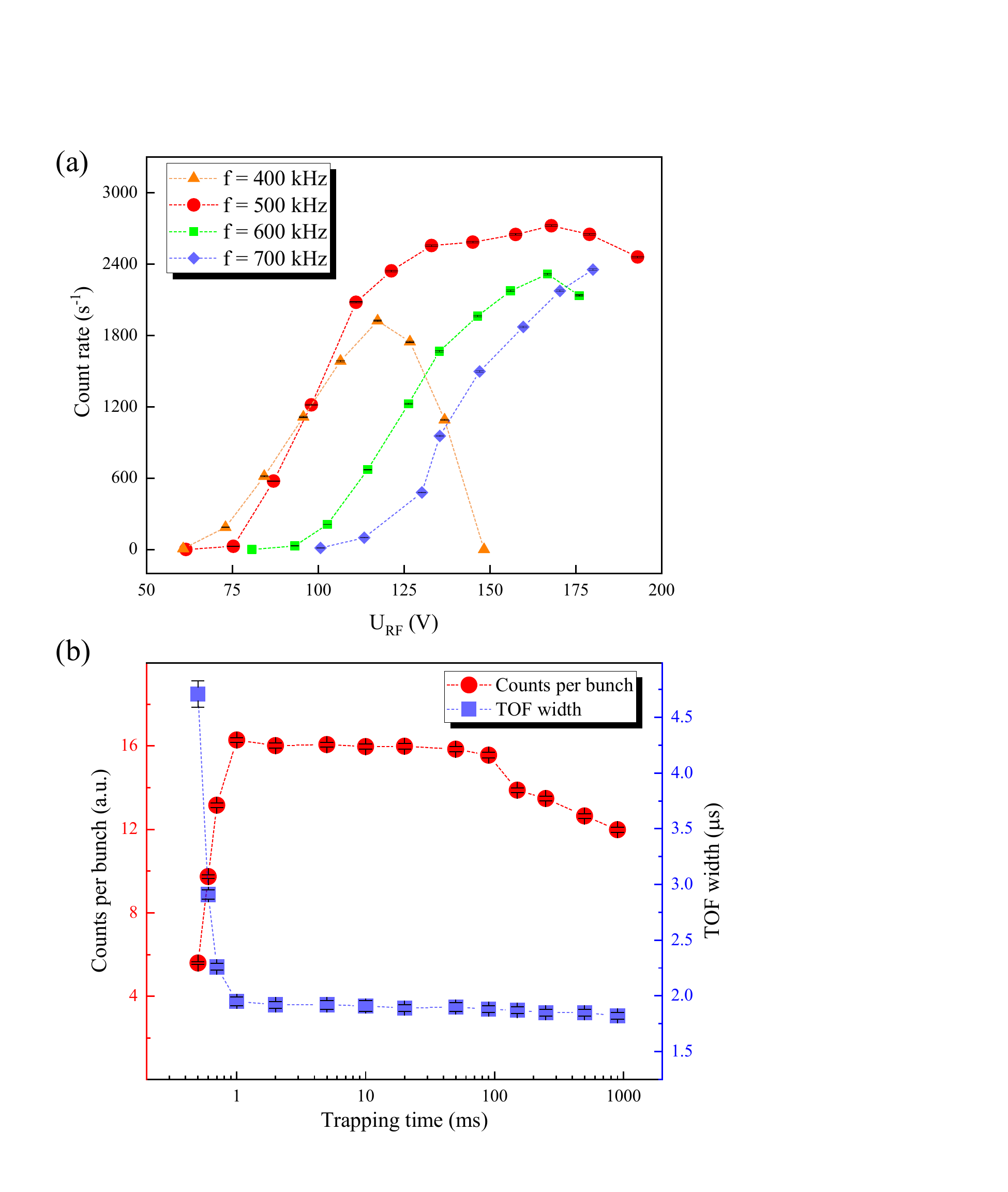} 
\end{center}
\vspace{-5mm}
\caption
{{(a) Count rate of the $^{85}$Rb ion bunches as a function of the RF voltage amplitude at different RF frequencies. 
(b) Counts per bunch and TOF width as functions of the RFQ trapping time.} }
\label{fig4}
\end{figure}

\begin{table*}
    \centering
    \caption{{A summary of the optimized operating parameters for the RFQ cooler-buncher system, tested with Rubidium ions from the surface ion source. 
    All values are directly read from the device, except for the DC potential gradient along the axial direction, which is calculated based on the voltages applied to the 12 DC electrodes.}}
    \label{tab1}
    \begin{tabular}{>{\centering\arraybackslash}p{70mm}|c|>{\centering\arraybackslash}p{70mm}|c}
         \hline
         Parameter&  Value&  Parameter& Value\\
         \hline
         Element&  Rb&  DC Potential Gradient (V/cm)& 1.28\\
         Ion Source Voltage HV$_{0}$ (V)&  29986.1&  Extraction Endcap (release) (V)& HV$_{1}-800$\\
         RFQ HV Platform Voltage HV$_{1}$ (V)&  29966.5&  Extraction Endcap (trapping) (V)& HV$_{1}+1600$\\
         Injection Cone (V) &  HV$_{1}-2400$&  Extraction Cone (V)& HV$_{1}-2000$\\
         Injection Endcap (V) &  HV$_{1}-13$&  RF Amplitude (V)& 167\\
         Trapping time (ms)&  5&  RF Frequency (kHz)& 500\\
         \hline
    \end{tabular}
\end{table*}

\begin{figure*}
\begin{center}
\includegraphics[width=0.7\linewidth]{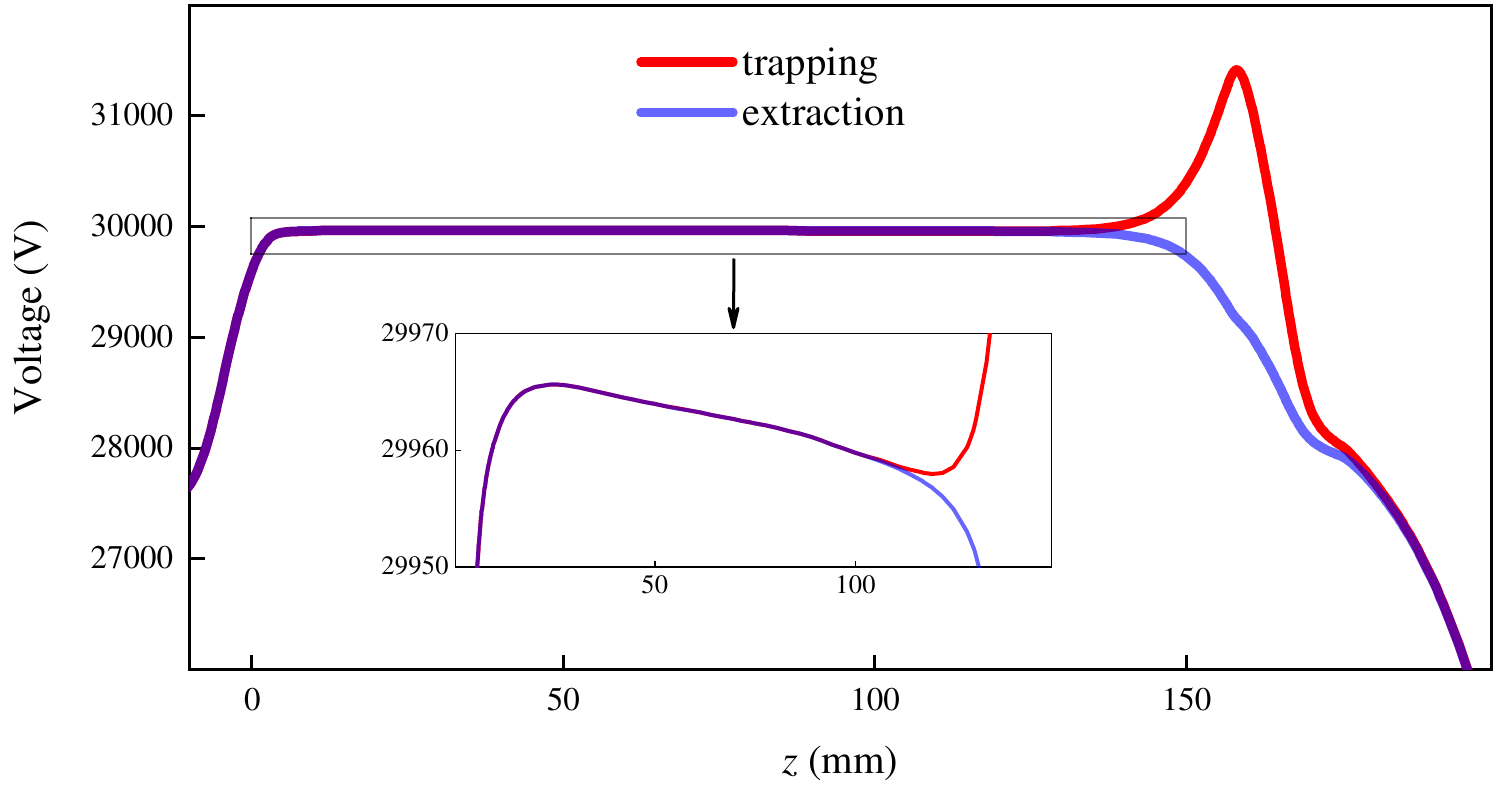} 
\end{center}
\vspace{-5mm}
\caption
{
The electrostatic potential along the RFQ axial direction calculated using voltages listed in Table~\ref{tab1}. The red line indicates the potential in the trapping mode, while the blue line corresponds to the extraction mode.} 
\label{fig-remix}
\end{figure*}

\begin{figure*}
\begin{center}
\includegraphics[width=0.9\textwidth]{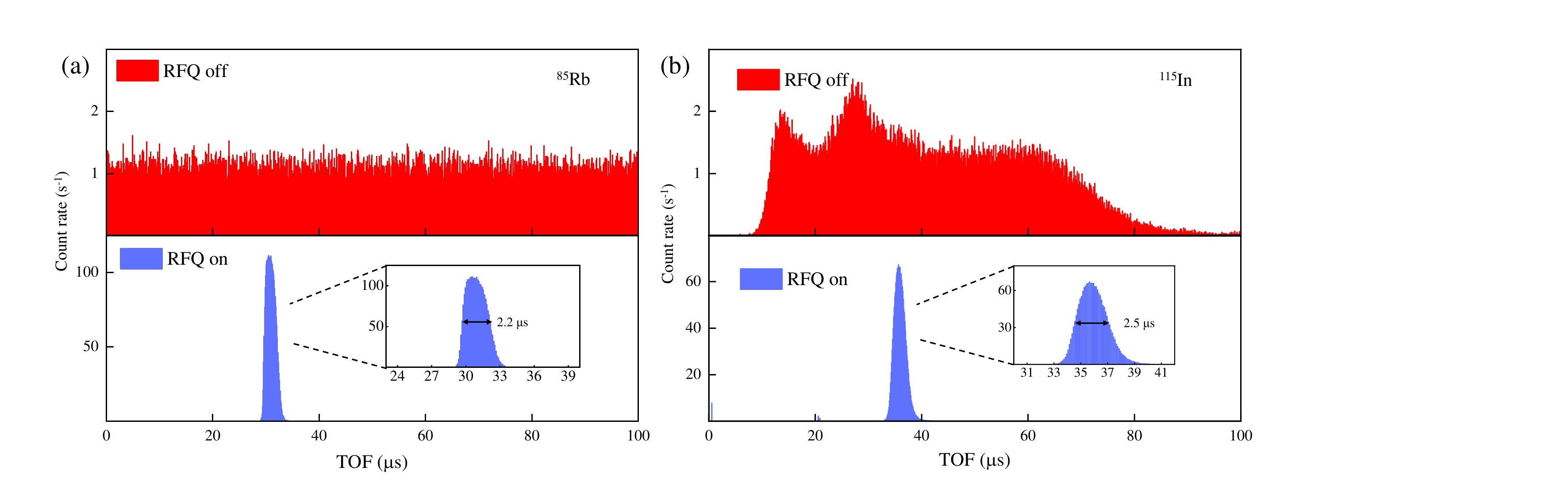} 
\end{center}
\vspace{-5mm}
\caption{{TOF spectra of stable $^{85}$Rb and $^{115}$In ions detected by the MagneTOF ion detector, with the RFQ cooler-buncher operation off and on. (a) TOF spectra of $^{85}$Rb produced from the surface ion source. (b) TOF spectra of $^{115}$In produced from the laser ablation ion source.}}
\label{fig5}
\end{figure*}

\begin{figure*}
\begin{center}
\includegraphics[width=1\textwidth]{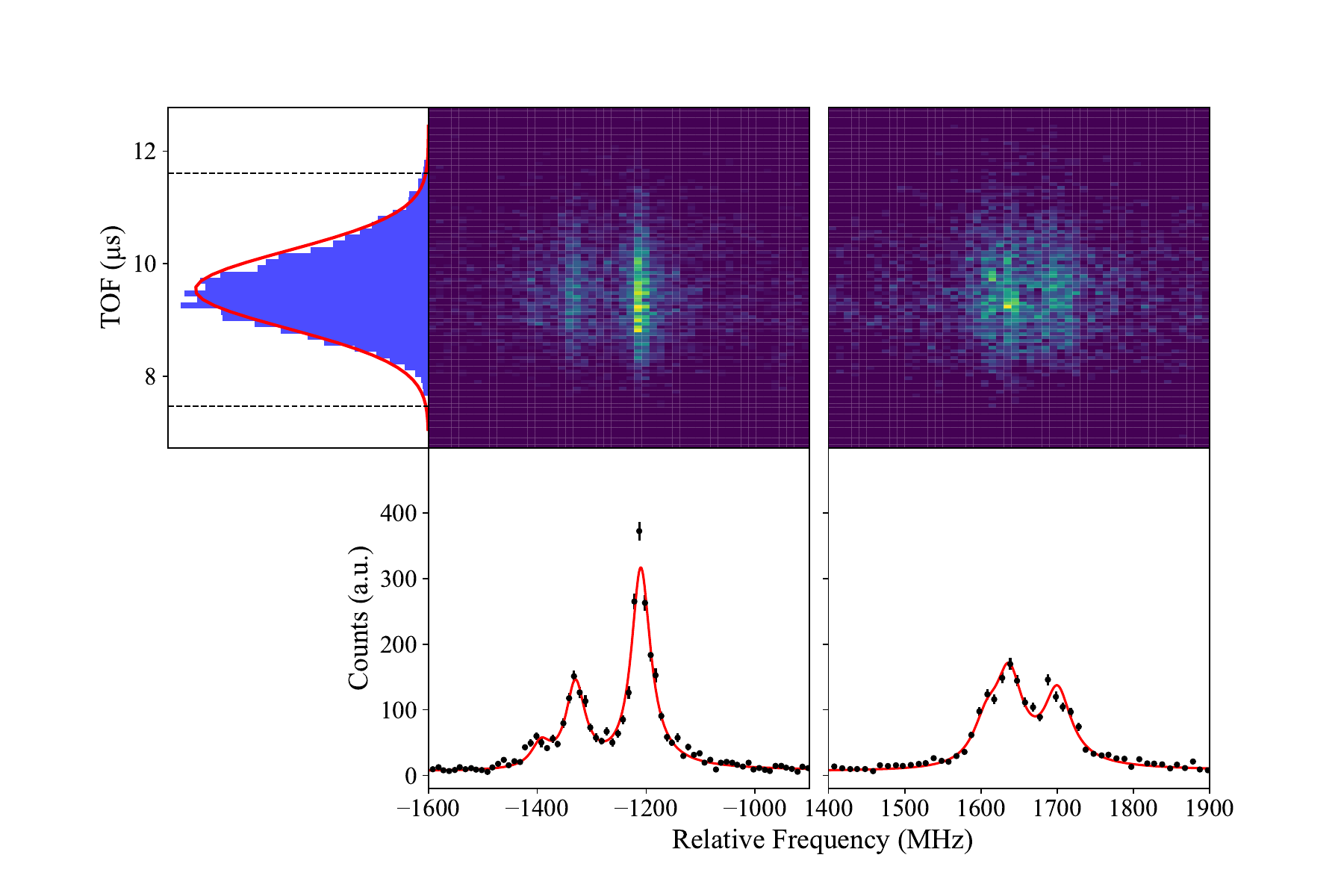} 
\end{center}
\vspace{-5mm}
\caption{The measured color-coded two-dimensional spectrum of TOF versus frequency for $^{85}$Rb (see Ref.~\cite{hu2025} for details). The projections onto the $x$-axis and $y$-axis correspond to the HFS spectrum (a) and TOF spectrum (b), respectively. }
\label{fig6}
\end{figure*}

The RFQ cooler-buncher system described above has been successfully manufactured and installed at Peking University. Since this system will be used to generate bunched ion beams for subsequent collinear resonance laser spectroscopy experiments, our primary focus is the TOF spectrum width of the ion bunches and overall transmission efficiency. In order to evaluate performance, the bunched ions are detected by a MagneTOF ion detector (ETP, 14924 MagneTOF$^{\rm{TM}}$ Mini) located downstream in the laser spectroscopy setup, as shown in Fig.~1 of Ref.~\cite{CRIS-NIMB-PKU}. The ion bunches generated from the cooler-buncher are delivered towards the ion detector through a 90 degree dipole magnet, as depicted in Fig.~\ref{fig2}. 

Given the large number of parameters involved in the RFQ cooler-buncher system operation, a systematic testing approach is required.
During commissioning, each component of the RFQ system is individually tested and optimized, followed by iterative fine-tuning of various device parameters to achieve global optimization. For example, as shown in Fig.~\ref{fig4}, the influence of the RF parameters and trapping time on the released ion counts and TOF width is examined. Using stable Rb ions from the surface ion source, the maximum count rate is achieved with an RF voltage of $U_{\rm{RF}} = 167$~V and an RF frequency of $f = 500$~kHz. 
With regard to trapping time, when ions are confined in the RFQ for less than 1~ms, the duration is insufficient for effective cooling and trapping, resulting in reduced counts per bunch and a broader TOF width. As the trapping time increases, the released counts per bunch reach a maximum and remain on a plateau within the range of 1–100~ms. Beyond 100~ms, a gradual decline in counts is observed, suggesting potential ion losses during prolonged confinement. As shown in Fig.~\ref{fig4}(b), except for trapping time below 1~ms, the TOF width of the ion bunch remains constant at approximately 2~$\mu$s. Taking both the counts per bunch and the TOF width into consideration, a trapping time of 5~ms is adopted during commissioning. The RFQ is operated at a repetition rate of 100~Hz to enable synchronization with the 100~Hz laser pulses used in the subsequent collinear resonance ionization spectroscopy measurement~\cite{hu2025}.

Other parameters can also influence the properties of the bunched ion beam. For example, the DC- and DC+ voltages applied to the extraction endcap significantly affect both the ion count rate and the TOF spectrum. The potential gradient across the 12 DC electrodes will change the profile of the TOF spectrum (\textit{e.g.}, symmetrical or asymmetrical). 
The buffer gas flow rate in the range of 10–30 SCCM has only a minor effect on both the ion count rate and TOF width.
In principle, a TOF spectrum with a width as narrow as few tens of nanoseconds can be achieved, as demonstrated at MIRACLS, but this comes at the cost of an increased energy spread and possibly a loss in overall efficiency. 
However, considering the commonly used TOF width of approximately 2~$\mu$s for collinear laser spectroscopy~\cite{79Zn,CRIS-Fr}, the optimized parameters listed in Table~\ref{tab1} are obtained under conditions that yield an overall RFQ transmission efficiency of 60\% for both continuous mode and bunched mode, as measured by using the FC1 and FC2.
The electrostatic potential along the axial direction of the RFQ cooler-buncher, calculated using the actual voltage settings listed in Table~\ref{tab1}, is shown in Fig.~\ref{fig-remix}.
Under these conditions, the observed TOF spectra of $^{85}$Rb ions from the surface ion source and $^{115}$In ions from a laser ablation ion source are shown in Fig.~\ref{fig5}. A comparison of the TOF spectra with and without RFQ cooler-buncher operation clearly demonstrates that this system effectively compresses the continuous ion beam and the bunched ion beam with a large temporal width of $\sim$100 $\mu$s, narrowing it into a significantly narrower ion bunches with a temporal width of $\sim$2 $\mu$s, while maintaining an overall efficiency greater than 60\%.
In addition, the RFQ cooler-buncher demonstrated consistent and reliable performance during both this test and the laser spectroscopy experiment~\cite{hu2025}, operating continuously for 8 to 10~hours per day while providing sustainable beam quality without noticeable degradation. This further confirms the long-term operational reliability of the RFQ system.

The goal is to integrate the RFQ cooler-buncher system into the collinear resonance laser spectroscopy setup~\cite{CRIS-NIMB-PKU}, aimed at measuring the nuclear properties of unstable nuclei far from $\beta$-stability on the nuclear chart. However, as discussed in Ref.~\cite{CRIS-NIMB-PKU}, the earlier test of collinear resonance ionization laser spectroscopy, without the RFQ cooler-buncher system, encountered a major issue: an unacceptable overall ion beam transmission efficiency of less than 40\%. This problem primarily arose from the distorted shape of the ion beam delivered into the CLS beamline, which significantly impacted the overlap between the ion beam and laser beams within the meter-long interaction region, thereby reducing the resonance ionization efficiency~\cite{PengZhangPhD}. With the successful integration of the RFQ cooler-buncher between the ion source and the laser spectroscopy setup, we have now achieved an overall ion beam transmission efficiency of $>$80\%, from FC2 to the end section of the CLS beamline, as shown in Fig.~1 of Ref.~\cite{CRIS-NIMB-PKU}. This demonstrates that, in addition to producing narrower ion bunches, the RFQ cooler-buncher system also improves the ion beam profile by reducing its transverse emittance, which is crucial for achieving optimal laser-ion overlapping during the CLS experiment~\cite{hu2025}. On this basis, we have measured the high-resolution and high efficiency HFS spectrum of stable $^{85}$Rb, as shown in Fig.~\ref{fig6} (see Ref.~\cite{hu2025} for details). These results confirm that the RFQ has successfully fulfilled its intended functions for laser spectroscopy experiments.

\section{Conclusion and Near term experiment plan}

In summary, an RFQ cooler-buncher system and its commissioning test are presented. The results demonstrate that this system successfully fulfills its intended functions of compressing the ion beam into short bunches and improving the ion beam profile. With an efficiency of RFQ cooler-buncher exceeding 60\%, ion beam bunches with a temporal width of $\sim$2 µs are achieved. Furthermore, the integration of this newly-developed system into the offline collinear laser spectroscopy setup has largely improved the overall transmission efficiency of the ion beam due to the enhanced beam profile with a lower transversal emittance.

Building on these commissioning results, the RFQ cooler-buncher system is firstly used for the offline test of a collinear resonance ionization laser spectroscopy setup~\cite{hu2025}. Soon it will be integrated into the online laser spectroscopy experiment of unstable nuclei at BRIF facility~\cite{BRIF} targeting Rb and Cs isotopes. Using a heavy-element target, neutron-rich $^{85-100}$Rb and $^{147-150}$Cs isotopes have already been produced at BRIF facility with a yield exceeding 100 particles per second. Given the relatively large energy spread of the ion beam at BRIF, approximately 21 eV for $^{38}$K as observed in an earlier experiment~\cite{WANG2022166622}, the installation of the RFQ cooler-buncher system will be crucial not only for producing bunched ion beams but also for reducing the energy spread of the radioactive ion beam. This will undoubtedly enhance the sensitivity and spectral resolution of the planned online collinear resonance ionization laser spectroscopy experiments. 

\bibliographystyle{elsarticle-num-names}
\bibliography{reference}

\end{document}